# Graphical Models for Game Theory

**Michael Kearns**[*]
Syntek Capital
New York, New York

**Michael L. Littman**
AT&T Labs–Research
Florham Park, New Jersey

**Satinder Singh**
Syntek Capital
New York, New York

## Abstract

We introduce a compact graph-theoretic representation for multi-party game theory. Our main result is a provably correct and efficient algorithm for computing approximate Nash equilibria in one-stage games represented by trees or sparse graphs.

## 1 INTRODUCTION

In most work on multi-player game theory, payoffs are represented in *tabular* form: if $n$ agents play a game in which each player has (say) two actions available, the game is given by $n$ matrices, each of size $2^n$, specifying the payoffs to each player under any possible combination of joint actions. For game-theoretic approaches to scale to large multi-agent systems, *compact* yet *general* representations must be explored, along with algorithms that can efficiently manipulate them[1].

In this work, we introduce *graphical models* for multi-player game theory, and give powerful algorithms for computing their Nash equilibria in certain cases. An $n$-player game is given by an undirected graph on $n$ vertices and a set of $n$ matrices. The interpretation is that the payoff to player $i$ is determined entirely by the actions of player $i$ and his neighbors in the graph, and thus the payoff matrix for player $i$ is indexed only by these players. We thus view the global $n$-player game as being composed of interacting local games, each involving (perhaps many) fewer players. Each player's action may have global impact, but it occurs through the propagation of local influences.

[1]For multi-*stage* games, there is a large literature on compact state-based representations for the different stages of the game, such as stochastic games or extensive form games (Owen 1995). Our focus is on representing *one*-stage, multi-*player* games.

There are many common settings in which such graphical models may naturally and succinctly capture the underlying game structure. The graph topology might model the physical distribution and interactions of agents: each salesperson is viewed as being involved in a local competition (game) with the salespeople in geographically neighboring regions. The graph may be used to represent organizational structure: low-level employees are engaged in a game with their immediate supervisors, who in turn are engaged in a game involving their direct reports and their own managers, and so on up to the CEO. The graph may coincide with the topology of a computer network, with each machine negotiating with its neighbors (to balance load, for instance).

There is a fruitful analogy between our setting and Bayesian networks. We propose a representation that is universal: any $n$-player game can be represented by choosing the complete graph and the original $n$-player matrices. However, significant representational benefits occur if the graph degree is small: if each player has at most $k \ll n$ neighbors, then each game matrix is exponential only in $k$ rather than $n$. The restriction to small degree seems insufficient to avoid the intractability of computing Nash equilibria. All of these properties hold for the problem of representing and computing conditional probabilities in a Bayes net. Thus, as with Bayes nets, we are driven to ask the natural *computational* question: for which classes of graphs can we give efficient (polynomial-time) algorithms for the computation of Nash equilibria?

Our main technical result is an algorithm for computing Nash equilibria when the underlying graph is a tree (or can be turned into a tree with few vertex mergings). This algorithm comes in two related but distinct forms. The first version involves an approximation step, and computes an approximation of *every* Nash equilibrium. (Note that there may be an exponential or infinite number of equilibria.) This algorithm runs in time polynomial in the size of the representation (the tree and the associated local game matrices), and constitutes one of the few known cases in which equilibria can be efficiently computed for a large class of general-sum, multi-player games. The second ver-sion of the algorithm runs in exponential time, but allows the *exact* computation of all Nash equilibria in a tree. In an upcoming paper (Littman et al. 2001), we describe a polynomial-time algorithm for the exact computation of a single Nash equilibrium in trees. Our algorithms require only local message-passing (and thus can be implemented by the players themselves in a distributed manner).

## 2  RELATED WORK

Algorithms for computing Nash equilibria are well-studied. McKelvey and McLennan (1996) survey a wide variety of algorithms covering 2- and $n$-player games; Nash equilibria and refinements; normal and extensive forms; computing either a sample equilibrium or exhaustive enumeration; and many other variations. They note that $n$-player games are computationally much harder than 2-player games, in many important ways. The survey discusses approximation techniques for finding equilibria in $n$-player games. Several of the methods described are not globally convergent, and hence do not guarantee an equilibrium. A method based on simplicial subdivision is described that converges to a point with equilibrium-like properties, but is not necessarily near an equilibrium or an approximate equilibrium. In contrast, for the restricted cases we consider, our algorithms provide running time and solution quality guarantees, even in the case of general-sum, $n$-player games.

Nash (1951), in the paper that introduces the notion of Nash equilibria, gives an example of a 3-player, finite-action game, and shows it has a unique Nash equilibria. Although all payoffs are rational numbers, Nash shows that the players' action probabilities at the equilibrium are irrational. This suggests that no finite algorithm that takes rational payoffs and transforms them using addition, subtraction, multiplication, and division will be able to compute exact equilibrium policies in general. Thus, the existence of an exact algorithm for finding equilibria in games with tree-structured interactions shows that these games are somewhat simpler than general $n$-player games. It also suggests that approximation algorithms are probably unavoidable for general $n$-player games.

Several authors have examined graphical representations of games. Koller and Milch (2001) describe an extension of influence diagrams to representing $n$-player games, and suggest the importance of exploiting graphical structure in solving normal-form games. La Mura (2000) describes a closely related representation, and provides globally convergent algorithms for finding Nash equilibria.

## 3  PRELIMINARIES

An $n$-player, two-action[2] game is defined by a set of $n$ matrices $M_i$ ($1 \leq i \leq n$), each with $n$ indices. The entry $M_i(x_1, \ldots, x_n) = M_i(\vec{x})$ specifies the payoff to player $i$ when the joint action of the $n$ players is $\vec{x} \in \{0, 1\}^n$ [3]. Thus, each $M_i$ has $2^n$ entries. If a game is given by simply listing the $2^n$ entries of each of the $n$ matrices, we will say that it is represented in *tabular* form.

The actions 0 and 1 are the *pure strategies* of each player, while a *mixed* strategy for player $i$ is given by the probability $p_i \in [0, 1]$ that the player will play 0. For any joint mixed strategy, given by a product distribution $\vec{p}$, we define the expected payoff to player $i$ as $M_i(\vec{p}) = \mathbf{E}_{\vec{x} \sim \vec{p}}[M_i(\vec{x})]$, where $\vec{x} \sim \vec{p}$ indicates that each $x_j$ is 0 with probability $p_j$ and 1 with probability $1 - p_j$.

We use $\vec{p}[i : p'_i]$ to denote the vector which is the same as $\vec{p}$ except in the $i$th component, where the value has been changed to $p'_i$. A *Nash equilibrium* for the game is a mixed strategy $\vec{p}$ such that for any player $i$, and for any value $p'_i \in [0, 1]$, $M_i(\vec{p}) \geq M_i(\vec{p}[i : p'_i])$. (We say that $p_i$ is a *best response* to $\vec{p}$.) In other words, no player can improve their expected payoff by deviating unilaterally from a Nash equilibrium. The classic theorem of Nash (1951) states that for any game, there exists a Nash equilibrium in the space of joint mixed strategies (product distributions).

We will also use the standard definition for approximate Nash equilibria. An $\epsilon$-*Nash equilibrium* is a mixed strategy $\vec{p}$ such that for any player $i$, and for any value $p'_i \in [0, 1]$, $M_i(\vec{p}) + \epsilon \geq M_i(\vec{p}[i : p'_i])$. (We say that $p_i$ is an $\epsilon$-*best response* to $\vec{p}$.) Thus, no player can improve their expected payoff by more than $\epsilon$ by deviating unilaterally from an approximate Nash equilibrium.

An $n$-player *graphical game* is a pair $(G, \mathcal{M})$, where $G$ is an undirected graph on $n$ vertices and $\mathcal{M}$ is a set of $n$ matrices $M_i$ ($1 \leq i \leq n$), called the *local game matrices*. Player $i$ is represented by a vertex labeled $i$ in $G$. We use $N_G(i) \subseteq \{1, \ldots, n\}$ to denote the set of *neighbors* of player $i$ in $G$—that is, those vertices $j$ such that the undirected edge $(i, j)$ appears in $G$. By convention, $N_G(i)$ always includes $i$ itself. The interpretation is that each player is in a game with only their neighbors in $G$. Thus, if $|N_G(i)| = k$, the matrix $M_i$ has $k$ indices, one for each player in $N_G(i)$, and if $\vec{x} \in [0, 1]^k$, $M_i(\vec{x})$ denotes the payoff to $i$ when his $k$ neighbors (which include himself) play $\vec{x}$. The expected payoff under a mixed strategy $\vec{p} \in [0, 1]^k$ is defined analogously. Note that in the two-action case, $M_i$ has $2^k$ entries, which may be considerably

---

[2] For simplicity, we describe our results for the two-action case. However, we later describe an efficient generalization of the approximation algorithm to multiple actions.

[3] For simplicity, we shall assume all payoffs are bounded in absolute value by 1, but all our results generalize (with a linear dependence on maximum payoff).

smaller than $2^n$.

Since we identify players with vertices in $G$, and since it will sometimes be easier to treat vertices symbolically (such as $U, V$ and $W$) rather than by integer indices, we also use $M_V$ to denote the local game matrix for the player identified with vertex $V$.

Note that our definitions are entirely representational, and alter nothing about the underlying game theory. Thus, every graphical game has a Nash equilibrium. Furthermore, every game can be trivially represented as a graphical game by choosing $G$ to be the complete graph, and letting the local game matrices be the original tabular form matrices. Indeed, in some cases, this may be the most compact graphical representation of the tabular game. However, exactly as for Bayesian networks and other graphical models for probabilistic inference, any time in which the local neighborhoods in $G$ can be bounded by $k \ll n$, exponential *space* savings accrue. Our main results identify graphical structures for which significant *computational* benefits may also be realized.

## 4 ABSTRACT TREE ALGORITHM

In this section, we give an abstract description of our algorithm for computing Nash equilibria in trees (see Figure 1). By "abstract", we mean that we will leave unspecified (for now) the representation of a certain data structure, and the implementation of a certain computational step. After proving the correctness of this abstract algorithm, in subsequent sections we will describe two instantiations of the missing details—yielding one algorithm that runs in polynomial time and computes approximations of all equilibria, and another algorithm that runs in exponential time and computes all exact equilibria.

If $G$ is a tree, we can orient this tree by choosing an arbitrary vertex to be the root. Any vertex on the path from a vertex to the root will be called *downstream* from that vertex, and any vertex on a path from a vertex to a leaf will be called *upstream* from that vertex. Thus, each vertex other than the root has exactly one downstream neighbor (child), and perhaps many upstream neighbors (parents). We use $\text{Up}_G(U)$ to denote the set of vertices in $G$ that are upstream from $U$, including $U$ by definition.

Suppose that $V$ is the child of $U$ in $G$. We let $G^U$ denote the the subgraph induced by the vertices in $\text{Up}_G(U)$. If $v \in [0, 1]$ is a mixed strategy for player (vertex) $V$, $\mathcal{M}_{V=v}^U$ will denote the subset of matrices of $\mathcal{M}$ corresponding to the vertices in $\text{Up}_G(U)$, with the modification that the game matrix $M_U$ is collapsed by one index by fixing $V = v$. We can think of a Nash equilibrium for the graphical game $(G^U, \mathcal{M}_{V=v}^U)$ as an equilibrium "upstream" from $U$ (inclusive), *given* that $V$ plays $v$.

Suppose some vertex $V$ has $k$ parents $U_1, \ldots, U_k$, and the single child $W$. We now describe the data structures sent from each $U_i$ to $V$, and in turn from $V$ to $W$, on the downstream pass of the algorithm.

Each parent $U_i$ will send to $V$ a binary-valued table $T(v, u_i)$. The table is indexed by the continuum of possible values for the mixed strategies $v \in [0, 1]$ of $V$ and $u_i \in [0, 1]$ of $U_i$. The semantics of this table will be as follows: for any pair $(v, u_i)$, $T(v, u_i)$ will be 1 if and only if there exists a Nash equilibrium for $(G^{U_i}, \mathcal{M}_{V=v}^{U_i})$ in which $U_i = u_i$. Note that we will slightly abuse notation by letting $T(v, u_i)$ refer both to the entire table sent from $U_i$ to $V$, and the particular value associated with the pair $(v, u_i)$, but the meaning will be clear from the context.

Since $v$ and $u_i$ are continuous variables, it is not obvious that the table $T(v, u_i)$ can be represented compactly, or even finitely, for arbitrary vertices in a tree. As indicated already, for now we will simply assume a finite representation, and show how this assumption can be met in two different ways in later sections.

The initialization of the downstream pass of the algorithm begins at the leaves of the tree, where the computation of the tables is straightforward. If $U$ is a leaf and $V$ its only child, then $T(v, u) = 1$ if and only if $U = u$ is a best response to $V = v$ (step 2c of Figure 1).

Assuming for induction that each $U_i$ sends the table $T(v, u_i)$ to $V$, we now describe how $V$ can compute the table $T(w, v)$ to pass to its child $W$ (step 2(d)ii of Figure 1). For each pair $(w, v)$, $T(w, v)$ is set to 1 if and only if there exists a vector of mixed strategies $\vec{u} = (u_1, \ldots, u_k)$ (called a *witness*) for the parents $\vec{U} = (U_1, \ldots, U_k)$ of $V$ such that

1. $T(v, u_i) = 1$ for all $1 \leq i \leq k$; and

2. $V = v$ is a best response to $\vec{U} = \vec{u}, W = w$.

Note that there may be more than one witness for $T(w, v) = 1$. In addition to computing the value $T(w, v)$ on the downstream pass of the algorithm, $V$ will also keep a list of the witnesses $\vec{u}$ for each pair $(w, v)$ for which $T(w, v) = 1$ (step 2(d)iiA of Figure 1). These witness lists will be used on the upstream pass. Again, it is not obvious how to implement the described computation of $T(w, v)$ and the witness lists, since $\vec{u}$ is continuous and universally quantified. For now, we assume this computation can be done, and describe two specific implementations later.

To see that the semantics of the tables are preserved by the abstract computation just described, suppose that this computation yields $T(w, v) = 1$ for some pair $(w, v)$, and let $\vec{u}$ be a witness for $T(w, v) = 1$. The fact that $T(v, u_i) = 1$ for all $i$ (condition 1 above) ensures by induction that if $V$ plays $v$, there are upstream Nash equilibria in which each $U_i = u_i$. Furthermore, $v$ is a best response to the local settings $U_1 = u_1, \ldots, U_k = u_k, W = w'$ (condition 2 above).

> Algorithm **TreeNash**
> Inputs: Graphical game $(G, \mathcal{M})$ in which $G$ is a tree.
> Output: A Nash equilibrium for $(G, \mathcal{M})$.
>
> 1. Compute a depth-first ordering of the vertices of $G$.
>
> 2. (**Downstream Pass**) For each vertex $V$ in depth-first ordering (starting at the leaves):
>
>    (a) Let vertex $W$ be the child of $V$ (or nil if $V$ is the root).
>    (b) Initialize $T(w, v)$ to be 0 and the witness list for $T(w, v)$ to be empty for all $w, v \in [0, 1]$.
>    (c) If $V$ is a leaf:
>        i. For all $w, v \in [0, 1]$, set $T(w, v)$ to be 1 if and only if $V = v$ is a best response to $W = w$ (as determined by the local game matrix $M_V$).
>    (d) Else ($V$ is an internal vertex):
>        i. Let $\vec{U} = (U_1, \ldots, U_k)$ be the parents of $V$; let $T(v, u_i)$ be the table passed from $U_i$ to $V$ on the downstream pass.
>        ii. For all $w, v \in [0, 1]$ and all joint mixed strategies $\vec{u} = (u_1, \ldots, u_k)$ for $\vec{U}$:
>            A. If $V = v$ is a best response to $W = w$ and $\vec{U} = \vec{u}$ (as determined by the local game matrix $M_V$), and $T(v, u_i) = 1$ for $i = 1, \cdots, k$, set $T(w, v)$ to be 1 and add $\vec{u}$ to the witness list for $T(w, v)$.
>    (e) Pass the table $T(w, v)$ from $V$ to $W$.
>
> 3. (**Upstream Pass**) For each vertex $V$ in reverse depth-first ordering (starting at the root):
>
>    (a) Let $\vec{U} = (U_1, \ldots, U_k)$ be the parents of $V$ (or the empty list if $V$ is a leaf); let $W$ be the child of $V$ (or nil if $V$ is the root), and $(w, v)$ the values passed from $W$ to $V$ on the upstream pass.
>    (b) Label $V$ with the value $v$.
>    (c) (Non-deterministically) Choose any witness $\vec{u}$ to $T(w, v) = 1$.
>    (d) For $i = 1, \ldots, k$, pass $(v, u_i)$ from $V$ to $U_i$.

Figure 1: Abstract algorithm **TreeNash** for computing Nash equilibria of tree graphical games. The description is incomplete, as it is not clear how to finitely represent the tables $T(\cdot, \cdot)$, or how to finitely implement step 2(d)ii. In Section 5, we show how to implement a modified version of the algorithm that computes approximate equilibria in polynomial time. In Section 6, we implement a modified version that computes exact equilibria in exponential time.

Therefore, we are in equilibrium upstream from $V$. On the other hand, if $T(w, v) = 0$ it is easy to see there can be no equilibrium in which $W = w, V = v$. Note that the existence of a Nash equilibrium guarantees that $T(w, v) = 1$ for at least one $(w, v)$ pair.

The downstream pass of the algorithm terminates at the root $Z$, which receives tables $T(z, y_i)$ from each parent $Y_i$. $Z$ simply computes a one-dimensional table $T(z)$ such that $T(z) = 1$ if and only if for some witness $\vec{y}$, $T(z, y_i) = 1$ for all $i$, and $z$ is a best response to $\vec{y}$.

The upstream pass begins by $Z$ choosing any $z$ for which $T(z) = 1$, choosing any witness $(y_1, \ldots, y_k)$ to $T(z) = 1$, and then passing both $z$ and $y_i$ to each parent $Y_i$. The interpretation is that $Z$ will play $z$, and is instructing $Y_i$ to play $y_i$. Inductively, if a vertex $V$ receives a value $v$ to play from its downstream neighbor $W$, and the value $w$ that $W$ will play, then it must be that $T(w, v) = 1$. So $V$ chooses a witness $\vec{u}$ to $T(w, v) = 1$, and passes each parent $U_i$ their value $u_i$ as well as $v$ (step 3 of Figure 1). Note that the semantics of $T(w, v) = 1$ ensure that $V = v$ is a best response to $\vec{U} = \vec{u}, W = w$.

We have left the choices of each witness in the upstream pass non-deterministic to emphasize that the tables and witness lists computed represent *all* the Nash equilibria. Of course, a random equilibrium can be chosen by making these choices random. We discuss the selection of equilibria with desired global properties in Section 7.

**Theorem 1** *Algorithm* **TreeNash** *computes a Nash equilibrium for the tree game* $(G, \mathcal{M})$. *Furthermore, the tables and witness lists computed by the algorithm represent all Nash equilibria of* $(G, \mathcal{M})$.

## 5 APPROXIMATION ALGORITHM

In this section, we describe an instantiation of the missing details of algorithm **TreeNash** that yields a polynomial-time algorithm for computing *approximate* Nash equilibria for the tree game $(G, \mathcal{M})$. The approximation can be made arbitrarily precise with greater computational effort.

Rather than playing an arbitrary mixed strategy in $[0, 1]$, each player will be constrained to play a *discretized* mixed strategy that is a multiple of $\tau$, for some $\tau$ to be determined by the analysis. Thus, player $i$ plays $q_i \in \{0, \tau, 2\tau, \ldots, 1\}$, and the joint strategy $\vec{q}$ falls on the discretized $\tau$-grid

$\{0, \tau, 2\tau, \ldots, 1\}^n$. In algorithm **TreeNash**, this will allow each table $T(v, u)$ (passed from vertex $U$ to child $V$) to be represented in discretized form as well: only the $1/\tau^2$ entries corresponding to the $\tau$-grid choices for $U$ and $V$ are stored, and all computations of best responses in the algorithm are modified to be approximate best responses. We return to the details of the approximate algorithm after establishing an appropriate value for the grid resolution $\tau$.

To determine an appropriate choice of $\tau$ (which in turn will determine the computational efficiency of the approximation algorithm), we first bound the loss in payoff to any player caused by moving from an arbitrary joint strategy $\vec{p}$ to the nearest strategy on the $\tau$-grid.

Fix any mixed strategy $\vec{p}$ for $(G, \mathcal{M})$ and any player index $i$, and let $|N_G(i)| = k$. We may write the expected payoff to $i$ under $\vec{p}$ as:

$$M_i(\vec{p}) = \sum_{\vec{x} \in \{0,1\}^k} \left( \prod_{j=1}^{k} \alpha_j(x_j) \right) M_i(\vec{x}), \quad (1)$$

where we simply define $\alpha_j(x_j) = (p_j)^{1-x_j}(1-p_j)^{x_j}$. Note that $\alpha_j(x_j) \in [0, 1]$ always.

We will need the following preliminary lemma.

**Lemma 2** *Let $\vec{p}, \vec{q} \in [0, 1]^k$ satisfy $|p_i - q_i| \leq \tau$ for all $1 \leq i \leq k$. Then provided $\tau \leq 2/(k \log^2(k/2))$,*

$$\left| \prod_{i=1}^{k} p_i - \prod_{i=1}^{k} q_i \right| \leq (2k \log k)\tau.$$

**Proof:** By induction on $k$. Assume without loss of generality that $k$ is a power of 2. The lemma clearly holds for $k = 2$. Now by induction:

$$\prod_{i=1}^{k} q_i = \left( \prod_{i=1}^{k/2} q_i \right) \left( \prod_{i=(k/2)+1}^{k} q_i \right)$$

$$\leq \left( \prod_{i=1}^{k/2} p_i + k(\log(k/2))\tau \right) \times$$

$$\left( \prod_{i=(k/2)+1}^{k} p_i + k(\log(k/2))\tau \right)$$

$$\leq \left( \prod_{i=1}^{k} p_i \right) + 2k(\log(k/2))\tau + (k(\log(k/2))\tau)^2$$

$$= \left( \prod_{i=1}^{k} p_i \right) + 2k(\log k - 1)\tau + (k(\log(k/2))\tau)^2.$$

The lemma holds if $-2k\tau + (k(\log(k/2))\tau)^2 \leq 0$. Solving for $\tau$ yields $\tau \leq 2/(k \log^2(k/2))$. $\square$

**Lemma 3** *Let the mixed strategies $\vec{p}, \vec{q}$ for $(G, \mathcal{M})$ satisfy $|p_i - q_i| \leq \tau$ for all $i$. Then provided $\tau \leq 2/(k \log^2(k/2))$,*

$$|M_i(\vec{p}) - M_i(\vec{q})| \leq 2^{k+1}(k \log(k))\tau.$$

**Proof:** Applying Lemma 2 to each term of Equation (1) yields

$$|M_i(\vec{p}) - M_i(\vec{q})|$$

$$\leq \sum_{\vec{x} \in \{0,1\}^k} \left| \prod_{j=1}^{k} \alpha_j(x_j) - \prod_{j=1}^{k} \beta_j(x_j) \right| |M_i(\vec{x})|$$

$$\leq \sum_{\vec{x} \in \{0,1\}^k} (2k \log(k))\tau \leq 2^{k+1}(k \log(k))\tau$$

where $\alpha_j(x_j) = (p_j)^{1-x_j}(1 - p_j)^{x_j}$, $\beta_j(x_j) = (q_j)^{1-x_j}(1 - q_j)^{x_j}$, and we have used $|M_i(\vec{x})| \leq 1$. $\square$

Lemma 3 bounds the loss suffered by any player in moving to the nearest joint strategy on the $\tau$-grid. However, we must still prove that Nash equilibria are approximately preserved:

**Lemma 4** *Let $\vec{p}$ be a Nash equilibrium for $(G, \mathcal{M})$, and let $\vec{q}$ be the nearest (in $L_1$ metric) mixed strategy on the $\tau$-grid. Then provided $\tau \leq 2/(k \log^2(k/2))$, $\vec{q}$ is a $2^{k+2}(k \log(k))\tau$-Nash equilibrium for $(G, \mathcal{M})$.*

**Proof:** Let $r_i$ be a best response for player $i$ to $\vec{q}$. We now bound the difference $M_i(\vec{q}[i : r_i]) - M_i(\vec{q}) \geq 0$, which is accomplished by maximizing $M_i(\vec{q}[i : r_i])$ and minimizing $M_i(\vec{q})$. By Lemma 3, we have

$$|M_i(\vec{q}[i : r_i]) - M_i(\vec{p}[i : r_i])| \leq 2^{k+1}(k \log(k))\tau.$$

Since $\vec{p}$ is an equilibrium, $M_i(\vec{p}) \geq M_i(\vec{p}[i : r_i])$. Thus,

$$M_i(\vec{q}[i : r_i]) \leq M_i(\vec{p}) + 2^{k+1}(k \log(k))\tau.$$

On the other hand, again by Lemma 3,

$$M_i(\vec{q}) \geq M_i(\vec{p}) - 2^{k+1}(k \log(k))\tau.$$

Thus, $M_i(\vec{q}[i : r_i]) - M_i(\vec{q}) \leq 2^{k+2}(k \log(k))\tau$. $\square$

Let us now choose $\tau$ to satisfy $2^{k+2}(k \log(k))\tau \leq \epsilon$ and $\tau \leq 2/(k \log^2(k/2))$ (which is the condition required by Lemma 3), or

$$\tau \leq \min(\epsilon/(2^{k+2}(k \log(k))), 2/(k \log^2(k/2))).$$

Lemma 4 finally establishes that by restricting play to the $\tau$-grid, we are ensured the existence of an $\epsilon$-Nash equilibrium. The important point is that $\tau$ needs to be exponentially small only in the *local neighborhood* size $k$, not the total number of players $n$.

It is now straightforward to describe the details of our approximate algorithm **ApproximateTreeNash**. This algorithm is identical to algorithm **TreeNash** with the following exceptions:

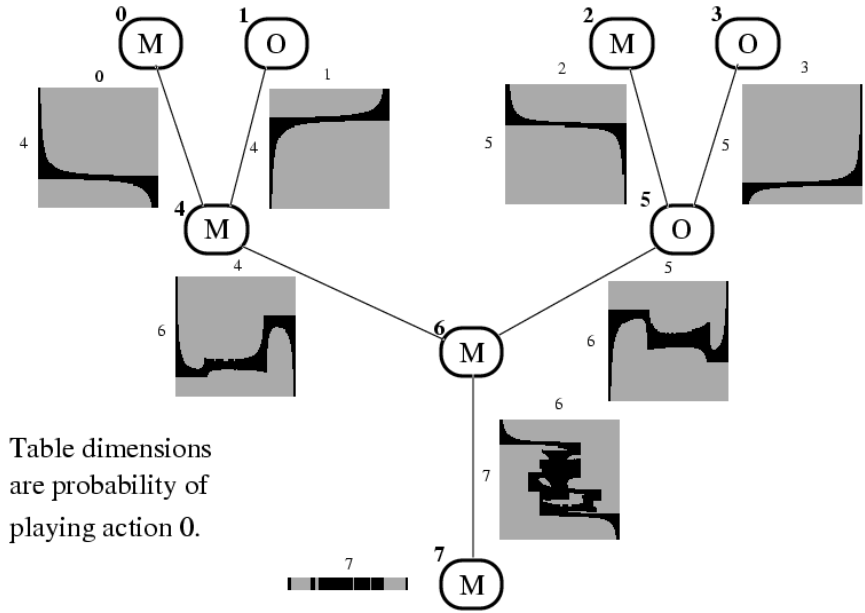

Figure 2: An example game, and the tables computed by the downstream pass of algorithm **ApproximateTreeNash**. Each vertex in the tree is a player with two actions. Although we omit the exact payoff matrices, intuitively each "M" player maximizes its payoff by matching its child's action, while each "O" player maximizes its payoff by choosing the opposite action of its child. The relative payoff for matching or unmatching is modulated by the parent values, and also varies from player to player within each vertex type. The grid figures next to each edge are a visual representation of the actual tables computed in the downstream pass of the algorithm, with the settings $\tau = 0.01$ and $\epsilon = 0.05$; 1s are drawn as black and 0s as gray. Approximate Nash equilibria for the game are computed from the tables by the upstream pass of the algorithm. One example of a pure equilibrium is $(0, 1, 1, 0, 0, 1, 0, 0)$; the tables represent a multitude of mixed equilibria as well.

- The algorithm now takes an additional input $\epsilon$.
- For any vertex $U$ with child $V$, the table $T(u, v)$ will contain only entries for $u$ and $v$ multiples of $\tau$.
- All computations of best responses in algorithm **TreeNash** become computations of $\epsilon$-best responses in algorithm **ApproximateTreeNash**.

Lemma 3 establishes that there will be such an approximate best response on the $\tau$-grid, while Lemma 4 ensures that the overall computation results in an $\epsilon$-Nash equilibrium. For the running time analysis, we simply note that each table has $(1/\tau)^2$ entries, and that the computation is dominated by the downstream calculation of the tables (Step 2(d)ii of algorithm **TreeNash**). This requires ranging over all table entries for all $k$ parents, a computation of order $((1/\tau)^2)^k$.

**Theorem 5** *For any $\epsilon > 0$, let*

$$\tau \leq \min(\epsilon/(2^{k+2}(k\log(k))), 2/(k\log^2(k/2))).$$

*Then **ApproximateTreeNash** computes an $\epsilon$-Nash equilibrium for the tree game $(G, \mathcal{M})$. Furthermore, for every exact Nash equilibrium, the tables and witness lists computed by the algorithm contain an $\epsilon$-Nash equilibrium that is within $\tau$ of this exact equilibrium (in $L_1$ norm). The running time of the algorithm is polynomial in $1/\epsilon$, $n$ and $2^k$, and thus polynomial in the size of $(G, \mathcal{M})$.*

See Figure 2 for an example of the behavior of algorithm **ApproximateTreeNash**.

## 6 EXACT ALGORITHM

In this section, we describe an implementation of the missing details of algorithm **TreeNash** that computes exact, rather than approximate, equilibria. In the worst case, the algorithm may run in time exponential in the number of vertices. We remind the reader that even this result is nontrivial, since there are no finite-time algorithms known for computing exact Nash equilibria in general-sum, multi-party games.

As before, let $\vec{U} = U_1, \ldots, U_k$ be the parents of $V$, and $W$ the child. We assume for induction that each table $T(v, u_i)$ passed from $U_i$ to $V$ on the downstream pass can be represented in a particular way—namely, that the set of $(v, u_i)$ pairs where $T(v, u_i) = 1$ is a finite union of axis-parallel rectangles (or line segments or points, degenerately) in the unit square. We formalize this representation by assuming each $T(v, u_i)$ is given by an ordered list called the $v$-list,

$$0 = v_1 \leq v_2 \leq \cdots \leq v_{m-1} \leq v_m = 1,$$

defining intervals of the mixed strategy $v$. For each $v$-interval $[v_\ell, v_{\ell+1}]$ $(1 \leq \ell \leq m)$, there is a subset of $[0, 1]$

$$I_1^{i,\ell} \cup \cdots \cup I_t^{i,\ell}$$

where each $I_j^\ell \subseteq [0,1]$ is an interval of $[0,1]$, and these intervals are disjoint without loss of generality. By taking the maximum, we can assume without loss of generality that the number of sets $t$ in the union associated with any $v$-interval is the same. The interpretation of this representation is that $T(v, u_i) = 1$ if and only if $v \in [v_\ell, v_{\ell+1}]$ implies $u_i \in I_1^{i,\ell} \cup \cdots \cup I_t^{i,\ell}$. We think of each $[v_\ell, v_{\ell+1}]$ as defining a horizontal strip of $T(v, u_i)$, while the associated union $I_1^{i,\ell} \cup \cdots \cup I_t^{i,\ell}$ defines vertical bars where the table is 1 within this strip.

We can assume that the tables $T(v, u_i)$ share a common $v$-list, by simply letting this common $v$-list be the merging of the $k$ separate $v$-lists. Applying algorithm **TreeNash** to this representation, we now must address the following question for the computation of $T(w, v)$ in the downstream pass. Fix a $v$-interval $[v_\ell, v_{\ell+1}]$. Fix any choice of $k$ indices $j_1, \ldots, j_k \in \{1, \ldots, t\}$. As we allow $\vec{u} = (u_1, \ldots, u_k)$ to range across the rectangular region $I_{j_1}^{1,\ell} \times \cdots \times I_{j_k}^{k,\ell}$, what is the set $\mathcal{W}$ of values of $w$ for which some $v \in [v_\ell, v_{\ell+1}]$ is a best response to $\vec{u}$ and $w$?

Assuming $v_\ell \neq 0$ and $v_{\ell+1} \neq 1$ (which is the more difficult case), a value in $[v_\ell, v_{\ell+1}]$ can be a best response to $\vec{u}$ and $w$ only if the payoff for $V = 0$ is identical to the payoff for $V = 1$, in which case *any* value in $[0,1]$ (and thus any value in $[v_\ell, v_{\ell+1}]$) is a best response. Thus, $T(w, v)$ will be 1 across the region $\mathcal{W} \times [v_\ell, v_{\ell+1}]$, and the union of all such subsets of $w \times v$ across all $m-1$ choices of the $v$-interval $[v_\ell, v_{\ell+1}]$, and all $t^k$ choices of the indices $j_1, \ldots, j_k \in \{1, \ldots, t\}$, completely defines where $T(w, v) = 1$. We now prove that for any fixed choice of $v$-interval and indices, the set $\mathcal{W}$ is actually a union of at most two intervals of $w$, allowing us to maintain the inductive hypothesis of finite union-of-rectangle representations.

**Lemma 6** *Let $V$ be a player in any $k+2$-player game against opponents $U_1, \ldots, U_k$ and $W$. Let $M_V(v, \vec{u}, w)$ denote the expected payoff to $V$ under the mixed strategies $V = v$, $\vec{U} = \vec{u}$, and $W = w$, and define $\Delta(\vec{u}, w) = M_V(0, \vec{u}, w) - M_V(1, \vec{u}, w)$. Let $I_1, \ldots, I_k$ each be continuous intervals in $[0,1]$, and let*

$$\mathcal{W} = \{w \in [0,1] : \exists \vec{u} \in I_1 \times \cdots \times I_k \;\; \Delta(\vec{u}, w) = 0\}.$$

*Then $\mathcal{W}$ is either empty, a continuous interval in $[0,1]$, or the union of two continuous intervals in $[0,1]$.*

**Proof:** We begin by writing

$$\Delta(\vec{u}, w) = \sum_{\vec{x} \in \{0,1\}^k, y \in \{0,1\}} (M_V(0, \vec{x}, y) - M_V(1, \vec{x}, y)) \times \left(w^{1-y}(1-w)^y \prod_{i=1}^k (u_i)^{1-x_i}(1-u_i)^{x_i}\right).$$

Note that for any $u_i$, $\Delta(\vec{u}, w)$ is a linear function of $u_i$, as each term of the sum above includes only either $u_i$ or $1 - u_i$.

Since $\Delta(\vec{u}, w)$ is a linear function of $u_i$, it is a monotonic function of $u_i$; we will use this property shortly.

Now by the continuity of $\Delta(\vec{u}, w)$ in $w$, $w \in \mathcal{W}$ if and only if $w \in \mathcal{W}_\geq \cap \mathcal{W}_\leq$, where

$$\mathcal{W}_\geq = \{w \in [0,1] : \exists \vec{u} \in I_1 \times \cdots \times I_k \;\; \Delta(\vec{u}, w) \geq 0\}$$

and

$$\mathcal{W}_\leq = \{w \in [0,1] : \exists \vec{u} \in I_1 \times \cdots \times I_k \;\; \Delta(\vec{u}, w) \leq 0\}.$$

First consider $\mathcal{W}_\geq$, as the argument for $\mathcal{W}_\leq$ is symmetric. Now $w \in \mathcal{W}_\geq$ if and only if $\max_{\vec{u} \in I_1 \times \cdots \times I_k} \{\Delta(\vec{u}, w)\} \geq 0$. But since $\Delta(\vec{u}, w)$ is a monotonic function of each $u_i$, this maximum occurs at one of the $2^k$ extremal points (vertices) of the region $I_1 \times \cdots \times I_k$. In other words, if we let $I_j = [\ell_j, r_j]$ and define the extremal set $E = \{\ell_1, r_1\} \times \cdots \times \{\ell_k, r_k\}$, we have

$$\mathcal{W}_\geq = \bigcup_{\vec{u} \in E} \{w : \Delta(\vec{u}, w) \geq 0\}.$$

For any fixed $\vec{u}$, the set $\{w : \Delta(\vec{u}, w) \geq 0\}$ is of the form $[0, x]$ or $[x, 1]$ by linearity, and so $\mathcal{W}_\geq$ (and $\mathcal{W}_\leq$ as well) is either empty, an interval, or the union of two intervals. The same statement holds for $\mathcal{W} = \mathcal{W}_\geq \cap \mathcal{W}_\leq$. Note that by the above arguments, $\mathcal{W}$ can be computed in time exponential in $k$ by exhaustive search over the extremal set $E$. □

Lemma 6 proves that any fixed choice of one rectangular region (where the table is 1) from each $T(v, u_i)$ leads to at most 2 rectangular regions where $T(w, v)$ is 1. It is also easy to show that the tables at the leaves have at most 3 rectangular regions. From this it is straightforward to show by induction that for any vertex $u$ in the tree with child $v$, the number of rectangular regions where $T(v, u) = 1$ is at most $2^{a_u} 3^{b_u}$, where $a_u$ and $b_u$ are the number of internal vertices and leaves, respectively, in the subtree rooted at $u$. This is a finite bound (which is at most $3^n$ at the root of the entire tree) on the number of rectangular regions required to represent any table in algorithm **TreeNash**. We thus have given an implementation of the downstream pass—except for the maintainence of the witness lists. Recall that in the approximation algorithm, we proved nothing special about the structure of witnesses, but the witness lists were finite (due to the discretization of mixed strategies). Here these lists may be infinite, and thus cannot be maintained explicitly on the downstream pass. However, it is not difficult to see that witnesses can easily be generated dynamically on the upstream pass (according to a chosen deterministic rule, randomly, non-deterministically, or with some additional bookkeeping, uniformly at random from the set of all equilibria). This is because given $(w, v)$ such that $T(w, v) = 1$, a witness is simply any $\vec{u}$ such that $T(v, u_i) = 1$ for all $i$.

Algorithm **ExactTreeNash** is simply the abstract algorithm **TreeNash** with the tables represented by unions of rectangles (and the associated implementations of computations

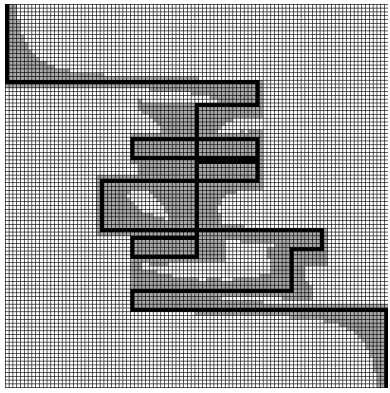

Figure 3: Example of a table produced by the exact algorithm. The table is the one generated for vertex 6 in Figure 2. Black cells indicate where the exact table is 1, while dark gray cells indicate where the approximate table is 1 for comparison. We see that the non-rectangular regions in Figure 2 are the result of the approximation scheme.

described in this section), and witnesses computed on the upstream pass. We thus have:

**Theorem 7** *Algorithm* **ExactTreeNash** *computes a Nash equilibrium for the tree game* $(G, \mathcal{M})$. *Furthermore, the tables computed by the algorithm represent all Nash equilibria of* $(G, \mathcal{M})$. *The algorithm runs in time exponential in the number of vertices of* $G$.

To provide a feel for the tables produced by the exact algorithm, Figure 3 shows the exact table for vertex 6 in the graph game in Figure 2.

## 7 EXTENSIONS

We have developed a number of extensions and generalizations of the results presented here. We describe some of them briefly, leaving details for the long version of this paper. At this writing, we have verified these extensions only for the approximation algorithm, and are working on the generalizations for the exact algorithm.

**Multiple Actions**. For ease of exposition, our approximation algorithm was presented for tree games in which players have only two actions available to them. By letting the tables $T(w, v)$ computed in the downstream pass of this algorithm be of the size necessary to represent the cross-product of the action spaces available to $V$ and $W$, we can recover the same result (Theorem 5) for the multiple-action case. The computational cost in the multiple-action case is exponential in the number of actions, but so is the size of the local game matrices (and hence the size of the representation of the tree game).

**Vertex Merging for Sparse Graphs**. The extension to multiple actions also permits the use of our approximation algorithm on arbitrary graphs. This is analogous to the use of the polytree algorithm on sparse, non-tree-structured Bayes nets. As in that case, the main step is the merging of vertices (whose action set will now be the product action space for the merged players) to convert arbitrary graphs into trees. To handle the merged vertices, we must ensure that the merged players are playing approximate best responses to each other, in addition to the upstream and downstream neighbors. With this additional bit of complexity (again proportional to the size of the representation of the final tree) we recover our result (Theorem 5).

As with the polytree algorithm, running time will scale exponentially with the largest number of merged players, so it is vital to minimize this cluster size. How best to accomplish this we leave to future work.

**Special Equilibria.** The approximation algorithm has the property that it finds an approximate Nash equilibrium for every exact Nash equilibrium. The potential multiplicity of Nash equilibria has led to a long line of research investigating Nash equilibria satisfying particular properties. By appropriately augmenting the tables computed in the downstream pass of our algorithm, it is possible to identify Nash equilibria that (approximately) maximize the following measures in the same time bounds:

- *Player Optimum*: Expected reward to a chosen player.
- *Social Optimum*: Total expected reward, summed over all players.
- *Welfare Optimum*: Expected reward to the player whose expected reward is smallest.

Equilibria with any of these properties are known to be NP-hard to find in the exact case, even in games with just two players (Gilboa and Zemel 1989).